\pdfoutput=1

\documentclass[8pt]{article}  \usepackage{times}
\usepackage{graphicx}

\usepackage{color}

\renewcommand{\baselinestretch}{1.0}

\begin{document}

\twocolumn[{\LARGE \textbf{Periodic solutions and refractory periods in the soliton theory for nerves and the locust femoral nerve\\*[0.0cm]}}

{\large Edgar Villagran Vargas$^{1,3,\ast}$, Andrei Ludu$^{1,4}$, Reinhold Hustert$^{5}$, Peter Gumrich$^{5}$, }\\
{\large Andrew D. Jackson$^2$, and Thomas Heimburg$^{1,\ast}$\\*[0.5cm]}
{\small {$^1$Membrane Biophysics Group, The Niels Bohr Institute, University of Copenhagen, Denmark}\\
\small {$^2$Niels Bohr International Academy, The Niels Bohr Institute, University of Copenhagen, Denmark}\\
\small {$^3$Departamento de F\'{\i}sica}, Universidad Aut\'{o}noma del Estado de M\'{e}xico, M\'{e}xico\\
\small {$^4$Department of Chemistry and Physics, Northwestern State University, Natchitoches, Louisiana}\\
\small {$^5$Institut f\"ur Zoologie und Anthropologie, Dept. Neurobiologie, University of G\"ottingen, Germany}\\}

{{\normalsize \textbf{ABSTRACT\hspace{0.5cm} Close to melting transitions it is possible to propagate solitary electromechanical pulses which reflect many of the experimental features of the nerve pulse including mechanical dislocations and reversible heat production. Here we show that one also obtains the possibility of periodic pulse generation when the boundary condition for the nerve is the conservation of the overall length of the nerve. This condition generates an undershoot beneath the baseline (`hyperpolarization') and a `refractory period', i.e., a minimum distance between pulses. In this paper, we outline the theory for periodic solutions to the wave equation and compare these results to action potentials from the femoral nerve of the locust ({\em locusta migratoria\/}).  In particular, we describe the frequently occurring minimum-distance doublet pulses seen in these neurons and compare them to the periodic pulse solutions.
}\\*[0.0cm] }}
]

\noindent\setlength{\parindent}{2cm}\footnotesize{\textbf{corresponding authors:} E. Villagran Vargas (villagran.e@gmail.com) and T. Heimburg (theimbu@nbi.dk).}\\
\noindent\setlength{\parindent}{2cm} \footnotesize{\textbf{keywords:} solitons, action potential, grasshopper, Hodgkin-Huxley model}\\
\noindent\setlength{\parindent}{2cm}\footnotesize{\textbf{abbreviations:} DPPC: 1,2-dipalmitoyl-sn-glycero-3-phosphocholine;}


\setlength{\parindent}{0cm}
\normalsize

\renewcommand{\baselinestretch}{1.0}\normalsize



\section*{Introduction}
In several recent publications we have proposed that the action potential in nerves is an electromechanical solitary wave or soliton \cite{Heimburg2005c, Heimburg2007b, Heimburg2008, Andersen2009}.  Although the propagation of mechanical pulses in nerves has been discussed before \cite{Wilke1912a, Wilke1912b, Cole1939, Hodgkin1945, Kaufmann1989e}, we have provided an quantitative formalism based on thermodynamics for how such pulses are generated and how they propagate. An important justification for the assumption of an electromechanical process is the experimental observation of reversible heat changes in phase with the action potential and a zero net heat release during the action potential (e.g., \cite{Abbott1958, Howarth1968, Howarth1975, Ritchie1985, Tasaki1989}). This finding is incompatible with an electrical picture of the nerve pulse (Hodgkin-Huxley model \cite{Hodgkin1952}) in which currents (of ions) passing through resistors (i.e., ion channel proteins) play a dominant role. The latter model would result in a net heat release that is not found in experiments.  The initial heat release during the action potential was also found to be much larger than the free energy required to charge the membrane capacitor, which rules out the possibility that a purely electrical picture is able to explain the nerve signal. The picture of the nerve pulse as an electromechanical phenomenon is further supported by the finding of mechanical changes of the nerve under the influence of the action potential, including changes in thickness and length (e.g., \cite{Wilke1912b, Iwasa1980b, Tasaki1989}).

The dynamic equation underlying the soliton description is based on the simple wave equation for sound with the lateral density of the nerve membranes as a variable. When describing sound propagation in air, it is usually assumed that the sound velocity, $c_0$, is constant.  This leads to the simple differential equation $\partial^2 \Delta \rho/\partial t^2=c_0^2 \partial^2 \Delta \rho/\partial x^2$. Close to the melting transition in membranes, however, the speed of sound is a sensitive function of density \cite{Heimburg1998, Halstenberg1998, Ebel2001, Schrader2002} and frequency \cite{Mitaku1982, Heimburg1998}. The simultaneous presence of non-linearity and dispersion gives rise to the possibility of localized excitations or solitons. The resulting differential equation is somewhat more sophisticated but now displays analytical localized solutions \cite{Lautrup2005, Heimburg2007b}.\footnote{Here, we use the term `soliton' as synonymous with `solitary wave'. Genuine solitons can pass through each other without dissipation, which is not the case for our pulses \cite{Lautrup2005}.  Strictly speaking, the term solitary wave is thus more appropriate.}  Due to the fact that biomembranes carry charges, these solitons are of an electromechanical or piezoelectric nature. Thus, they are not only localized voltage pulses but also display reversible mechanical features and a reversible change in heat that are in agreement with experiment. Further, since our theory is of a thermodynamic nature, it implicitly has a role for all thermodynamic variables, e.g., temperature, pressure, electric potential and the chemical potentials of drugs and ions. This implies that the generation of the nerve pulse can be affected by pH \cite{Heimburg2007c}, calcium concentration, and general or local anesthetics \cite{Heimburg2007b, Heimburg2007c}.

While a reversible electromechanical picture of the nerve pulse is compelling and able to address many more physical features of the nervous impulse than traditional models, there remain issues that have not yet been addressed. In particular, this includes refractory periods, hyperpolarization, and pulse trains. The refractory period is the minimum possible time between two nerve pulses. In the Hodgkin-Huxley picture it is a consequence of the complicated voltage and time dependence of the opening characteristics of ion channels.  In simplified terms, the refractory period is due to the relaxation time of the ion channel proteins after firing and is several milliseconds long, i.e., several pulse widths.  Hyperpolarization is the undershoot of the voltage below the resting potential during one phase of the action potential.

In the soliton model the action potential consists of a local region of the nerve with high chain order and high local density. An immediate consequence is that the excited nerve contracts. This has been observed in isolated nerves \cite{Wilke1912b, Tasaki1989}. However, it is difficult to imagine that nerves also contract in the native situation when they are embedded in tissue where distances are kept constant. Here, we demonstrate that one can obtain solutions for the electromechanical differential equation that maintain the length of the nerve. They consist of periodic pulses with an intrinsic minimal distance between pulses that is of the order of about 5-10 pulse widths. Using locust femoral nerves, we demonstrate that one finds minimal distances of exactly that order. In particular, we describe a phenomenon that we called pulse ``doublets" that displays a constant distance between pulses even if measured under different conditions. We have found such doublets in locust femoral nerves and also recently doublets and triplets in crayfish motor neurons. We show that the refractory period in the soliton model is a consequence of the conservation of mass and overall length of the nerve. This boundary condition also leads to periodic solutions or pulse trains and to hyperpolarization. 


\section*{Materials and Methods}
\textbf{Locust nerve recordings:}
Adult male and female {\em Locusta migratoria\/} were reared in a laboratory colony at about 30$^{\circ}$\,C and with a 12:12 hours dark-light cycle. Experiments were performed using amputated metathoracic legs which were mounted on a platform and fixed on its external lateral face with  PlasticineTM  (Modeling Clay KR 25, non-toxic, Becher G\"ottingen). The axis of rotation of the femoro-tibial joint was aligned with the origin, of a semicircular angular plane. 
Inside a Faraday cage and with the aid of an optical microscope, a small window of the postero-lateral cuticle was removed in order  to expose the peripheral lateral internal nerve (properly called posterior tegumental femoral nerve n5B2c \cite{Mucke1991, Hustert1999}, and the tibia was cut to a stump to allow oxygenation inside the leg via the trachea.  See figure \ref{Figure_Exp0} (bottom).  In order to record the electrical activity from the single antero-lateral (RDAL) and the two postero-lateral multipolar sensory afferents (RDPL and RVPL) (nomenclature according to \cite{Coillot1969}), monopolar platinum wire electrodes, 50\,$\mu$m in diameter, were hooked on the lateral internal  nerve n5b2c.

With the aid of a pipette, a small drop of locust saline (NACl 140\,mM, KCl 10\,mM, Na H$_2$PO$_4$ 5\,mM, CaCl$_2$ 2\,mM,Saccharose 90\,mM; pH 6.8) was deposited on the wounds to avoid the coagulation of the hemolymphe  and also used as a reference. A small amount of Vaseline (DAB 10, Roth GmbH, Karlsruhe, Germany) was added to isolate the nerve-electrode system as well as to avoid desiccation of the preparation.

The femoro-tibial angle was extended and fixed with an entomological pin taking care to avoid lateral strain on the joint. Although primitive, the use of a pin to fix the angle is the most precise method, because the tibia tends always to the flexion and this counterforce allows a perfect static equilibrium at the desired angle of extension.
All experiments were performed at room temperature of 22--24$^{\circ}$\,C.

\textbf{Detailed Procedure:}

Results presented in this report belong to 10 samples taken from different animals.  In all cases, three electrodes were hooked on the lateral internal nerve n5b2c, and a small amount of Vaseline was added to isolate each electrode-nerve subsystem. The distance from the axis of rotation of the joint, O, to the electrodes was 10, 12 and 15\,mm, respectively. We did not put the first electrode closer to the origin of the system for fear of damaging the receptors, and the last electrode could not be hooked beyond 17\,mm because the nerve under study joins the nerve n5b2 at that point.  In all experiments, the tibia was extended from 0 to 80$^{\circ}$ to verify that no activity could be recorded. Afterwards, at each angle from 80  to 150$^{\circ}$ in steps of 10$^{\circ}$,  the tibia was fixed completely with an entomological pin and sessions of three minutes were recorded for each angle. We found that 150$^{\circ}$ is the limit of extension since beyond this angle the end of the tibia causes pressure against the distal dorsal edge of the femur and therefore the conditions are no longer physiologic \cite{Coillot1969}. All experiments were performed at room temperature of 22--24$^{\circ}$\,C.

The action potentials were picked up externally by platinum wire hook electrodes. A preamplifier (four channels, 1000x amplification, no pass filter, 9V battery source, workshop of the Institute for Zoology, University of G\"ottingen) was used for primary signal amplification within the Faraday cage, and secondary amplifier (four channels, 10$\times$ amplification, 100\,Hz high pass filter, 30\,kH low  pass filter, 15V transformer source) used to pre-filter the signals. The amplified action potentials were digitized via PC based data acquisition hardware (Packard Bell desktop computer, 8 channel RUN Technologies acquisition port, PCI-DAS1200Jr board, Microsoft Windows XP Professional SP2, 2.66GHz Pentium(R) 4 CPU, 512 MB RAM) and recorded and analyzed with the DATAPAC 2K2 software (RUN Technologies, Mission Viejo, CA, USA).

\textbf{Analysis of the recordings:}

The smallest digital recording sample period achieved on the system was 49.6\,$\mu$s; periods of 89.6 and 148.8\,$\mu$s were chosen for longer recordings. No software-based secondary filtering was performed on the recorded channels.  Pulse height was defined by the voltage distance between maximum depolarization and maximum hyperpolarization in order to account for a potential signal baseline slope during the recording. Pulse duration was defined as the time difference between maximum depolarization and maximum hyperpolarization during a pulse. In all pulse trains, the onset of the individual signals was defined by the maximum depolarization within the pulse, the offset by the maximum hyperpolarization. Signal velocity along the nerve was estimated by subtracting the onset times of time-linked signals between the channels and multiplying the inverse by the measured distance between the corresponding hook electrodes. The instantaneous pulse period was defined as the period between the current onset and the onset of the next signal within a pulse train.

In each region of interest, pulses (APs) were first identified and time localized via a simple threshold pass analysis. The threshold was established visually, positioned at roughly 80\% of the maximum depolarization amplitude of the smallest unit of interest, then adjusted manually in such a way as to maximize the sensitivity and specificity regarding AP localization, ideally with no noise spikes falsely identified as APs. The onset of the marked pulses was then moved to the first peak depolarization, the offset to the first peak hyperpolarization.


\section*{Theory}
In the following we want to derive the equations of motion in a cylindrical membrane along the coordinate $x$. The lateral density, $\rho^A$, of a membrane is larger in the gel phase of the membrane than in the fluid phase because the area per lipid molecule changes by about 24\%.  For this reason, one can induce a phase transition not only by adjusting the temperature but also by changes in lateral pressure or other thermodynamic variables.  

Let us consider the nerve axon as a one-dimensional cylinder with lateral density excitations moving along the coordinate $x$.  As outlined in \cite{Heimburg2005c}, sound propagation in the absence of dispersion is governed by the equation
\begin{equation}
    \frac{\partial^2}{\partial t^2} \Delta \rho^A =
    \frac{\partial}{\partial x} \left(c^2 \frac{\partial}{\partial x}\Delta \rho^A\right) \ ,
    \label{eq:Theory1}
\end{equation}
where $\Delta \rho^A = \rho^A - \rho_{0}^A$ is the change in the area density of the membrane as a function of $x$ and $t$ and $c=\sqrt{1/(\kappa_S^A \rho^A)}$ is the density-dependent velocity of sound.  Here, $\rho_0^A$ is the density of the membrane at physiological conditions and is slightly above the melting transition. The above equation is related to the Euler equation in hydrodynamics and will not be derived here.  To the extent that the compressibility is independent of the density and the amplitude of the propagating density wave, $\Delta \rho^A << \rho_{0}^A$, the sound velocity is approximately constant ($c=c_0$). Eq.(\ref{eq:Theory1}) reduces to
\begin{equation}
    \frac{\partial^2}{\partial t^2} \Delta \rho^A =
    c_{0}^2 \frac{\partial^2}{\partial x^2} \Delta \rho^A \ .
    \label{eq:Theory2}
\end{equation}
This is the well-known equation that governs sound propagation in air. However, in several publications we showed that the compressibility, $\kappa_{S}^A$, depends sensitively on temperature (and therefore also on density) close to the melting transition of lipid membranes \cite{Heimburg1998, Halstenberg1998, Ebel2001, Schrader2002}. Therefore, the above simplification cannot be made.  Both the liquid and gel phases are relatively incompressible.  At densities near the phase transition where the two phases co-exist, a small increase in pressure can cause a significant increase in density by converting liquid to gel.  Near this phase transition, the compression modulus is dramatically smaller.  We thus approximate the sound velocity, $c$, as 
\begin{equation}
    c^2=\frac{1}{\rho^A \kappa_{S}^A}=c_{0}^2+p\Delta\rho^A 
    +q(\Delta\rho^A)^2
    \label{eq:Theory3}
\end{equation}
with $p < 0$ and $q > 0$.  The sound velocity is also frequency dependent \cite{Mitaku1982}. This means that the system is dispersive, which is a necessary requirement for soliton formation.  At very low frequencies, $\omega \approx 0$, the adiabatic compressibility approaches the isothermal compressibility.

In the following, we use the values for the isothermal compressibility as a low frequency approximation of the the adiabatic compressibility. For unilamellar DPPC vesicles we found $c_{0}=176.6$\,m/s, $p = -16.6 \, c_{0}^2/\rho_0^A$ and $q = 79.5 \, c_{0}^2 /(\rho_{0}^A)^2$ with $\rho_{0}^A = 4.035 \times 10^{-3}$\,g/m$^2$, assuming a bulk temperature of T=45$^{\circ}$C \cite{Heimburg2005c}.  Similar values are obtained for biological membranes at physiological temperature. 

The isentropic compressibility decreases at higher frequencies and leads to a propagation velocity which increases with increasing frequency.  We approximate such dispersive effects by introducing a term, $-h \partial^4 \Delta \rho^A/\partial x^4$ with $h > 0$, in eq.\,(\ref{eq:Theory1}).  As demonstrated in \cite{Heimburg2005c}, this term implies a linear dependence of the pulse propagation velocity on frequency.  This form of the dispersion term represents the leading term in a typical long-wavelength dispersion relation.  In the absence of quantitative knowledge of the dispersion present in nerves, this term has the advantage of being relatively passive in that changes in $h$ alter the spatial size of solitary waves but not their functional form.  The inclusion of additional terms can lead to soliton instability and should not be considered without empirical justification for the form chosen.

The equation governing sound propagation now becomes
\begin{eqnarray}
    \frac{\partial^2}{\partial t^2} \Delta \rho^A & = &
    \frac{\partial}{\partial x} \left[\left(c_0^2 + p \Delta \rho^A + q
    (\Delta \rho^A)^2 \right) \frac{\partial}{\partial x} \Delta 
    \rho^A\right] \nonumber\\
    && - h
    \frac{\partial^4}{\partial x^4} \Delta \rho^A \ .
    \label{eq:Theory4}
\end{eqnarray}
\begin{figure*}[htb!]
    \begin{center}
	\includegraphics[width=15cm]{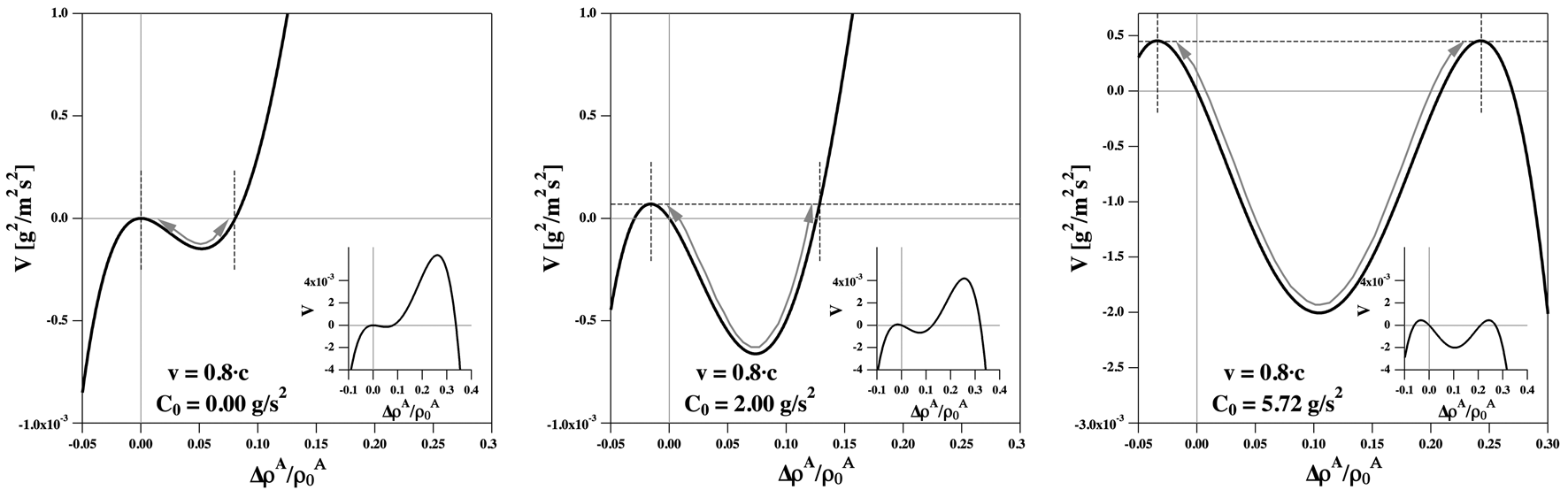}
	\parbox[c]{17cm}{ \caption{\textit{The potential $V(\Delta \rho^A)$ from eq.\,(\ref{eq:Theory8}) for three different integration constants, $C$, and a velocity $v=0.8\cdot  c_0$. The inserts show the potentials in a larger density range showing that they all possess two maxima and one minimum. The electromechanical pulses correspond to a movement in the potential between the limits given by the two dashed lines. Left: For $C=0$ g/cm$^2$ one finds a singular solution given in Fig.\,\ref{Figure_Theor2}, right top. This solution displays a baseline of $\Delta \rho^A=0$. Center: For $C=2$ g/s$^2$ one obtains a periodic solution Fig.\,\ref{Figure_Theor2}, right, 3rd trace from top. This  solution displays a baseline of $\Delta \rho^A<0$.  Right: For $C=5.72$g/s$^2$ one obtains a potential with two maxima of equal height, corresponding to the periodic solution given   in Fig.\,\ref{Figure_Theor2}, right bottom. This solution also displays a baseline of $\Delta \rho^A<0$, and the peak hits a maximum and broadens. }
	\label{Figure_Theor1}}}
    \end{center}
\end{figure*}
This equation is closely related to the Boussinesq equation \cite{Remoissenet1999}.  Since we are interested in finding solutions that propagate without changing their shape, we transform to a moving coordinate system with $z=x-vt$. The time- and space-dependent differential equation (eq. \ref{eq:Theory4}) can be transformed into a time-independent equation by using $\partial \Delta \rho/\partial x=(\partial \Delta \rho/\partial z)\cdot(\partial \Delta z/\partial x)=\partial \Delta \rho/\partial z$ and  $\partial \Delta \rho/\partial t=(\partial \Delta \rho/\partial z)\cdot(\partial z/\partial t)=-v \partial \Delta \rho/\partial z$.

Eq.(\ref{eq:Theory4}) can now be rewritten as
\begin{eqnarray}
    v^2\frac{\partial^2}{\partial z^2} \Delta \rho^A & = &
    \frac{\partial}{\partial z} \left[\left(c_0^2 + p \Delta \rho^A + q
    (\Delta \rho^A)^2 \right) \frac{\partial}{\partial z}\Delta \rho^A\right] \rightarrow\nonumber\\
    &\rightarrow& - h
    \frac{\partial^4}{\partial z^4} \Delta \rho^A \ .
    \label{eq:Theory5}
\end{eqnarray}
We now search for periodic solutions. We can now perform two integrals with respect to $z$ from one minimum of the periodic solution to the next
\begin{equation}
h \frac{\partial^2}{\partial z^2} \Delta \rho^A = (c_0^2 - v^2 )\Delta \rho^A + \frac{1}{2}p (\Delta \rho^A)^2 + \frac{1}{3} q(\Delta
\rho^A)^3 + C\ .
\label{eq:Theory6}
\end{equation}
This can be easily checked, since the second derivative of eq.\,(\ref{eq:Theory6}) yields eq.\,(\ref{eq:Theory5}). $C$ is an integration constant. The second integration constant disappears for reasons of symmetry.

Now multiply both sides of eq.\,(\ref{eq:Theory6}) by $\partial (\Delta \rho^A)/\partial z$ and integrate once more to yield
\begin{eqnarray}
    h\,\left( \frac{\partial}{\partial z}\Delta \rho^A \right)^2&=&(c_0^2 - v^2 ) \, (\Delta \rho^A)^2 + \frac{p}{3} \, (\Delta \rho^A)^3 \nonumber\\
    && + \frac{q}
    {6} \, (\Delta \rho^A)^4  +C\Delta \rho + V_0 \ .
    \label{eq:Theory7}
\end{eqnarray}
Here, $V_0$ is a further integration constant. In this equation there is one term that is proportional to $(\partial \Delta \rho^A/\partial z)^2$ and a second term that explicitly only depends on $\Delta \rho^A$.  Since a mechanical analogy is useful, we call the first term a ``kinetic term" and the second one a ``potential term":
\begin{equation}
    \underbrace{h\,\left( \frac{\partial}{\partial z}\Delta \rho^A \right)^2}_{\mbox{\scriptsize`kinetic term'}} - \rightarrow
    \label{eq:Theory8}
\end{equation}
\begin{displaymath}
\underbrace{-\left[(c_0^2 - 
    v^2 ) \, (\Delta \rho^A)^2 - \frac{p}{3} \, (\Delta \rho^A)^3 - \frac{q}{6} \, (\Delta \rho^A)^4 -C\Delta \rho\right]}_{\mbox{\scriptsize 
    `potential term'} \equiv V(\Delta \rho^A)} = V_0 \ .
\end{displaymath}
or
\begin{equation}
    h\,\left( \frac{\partial}{\partial z}\Delta \rho^A \right)^2 +V(\Delta \rho^A)= V_0 \ .
    \label{eq:Theory9}
\end{equation}
This equation is formally equivalent to the equation of motion $\frac{1}{2} m v^2 + V(x)=\mbox{const.}$ familiar from classical mechanics where $z$ plays the role of time and the density change $\Delta \rho^A$ serves as a spatial coordinate.  The shape of the soliton is a result of the motion in this potential, and all solutions of eq.\,(\ref{eq:Theory9}) can be classified according to this potential energy surface \cite{Eichmann2002}.  It should be stressed that this analogy to classical mechanics must not be taken too literally. It is just an analogy because one of the terms depends on the square of the derivative of the density with respect to the coordinate, while the second term is an analytic function of the density.  $V_0$ is not relevant because it moves the potential up or down without any influence on the solutions of the differential equation. The value for the integration constant $C$ will depend on experimental boundary conditions.

In the following we explore the solutions of eq\,(\ref{eq:Theory9}).


\begin{figure*}[htb!]
    \begin{center}
	\includegraphics[width=15cm]{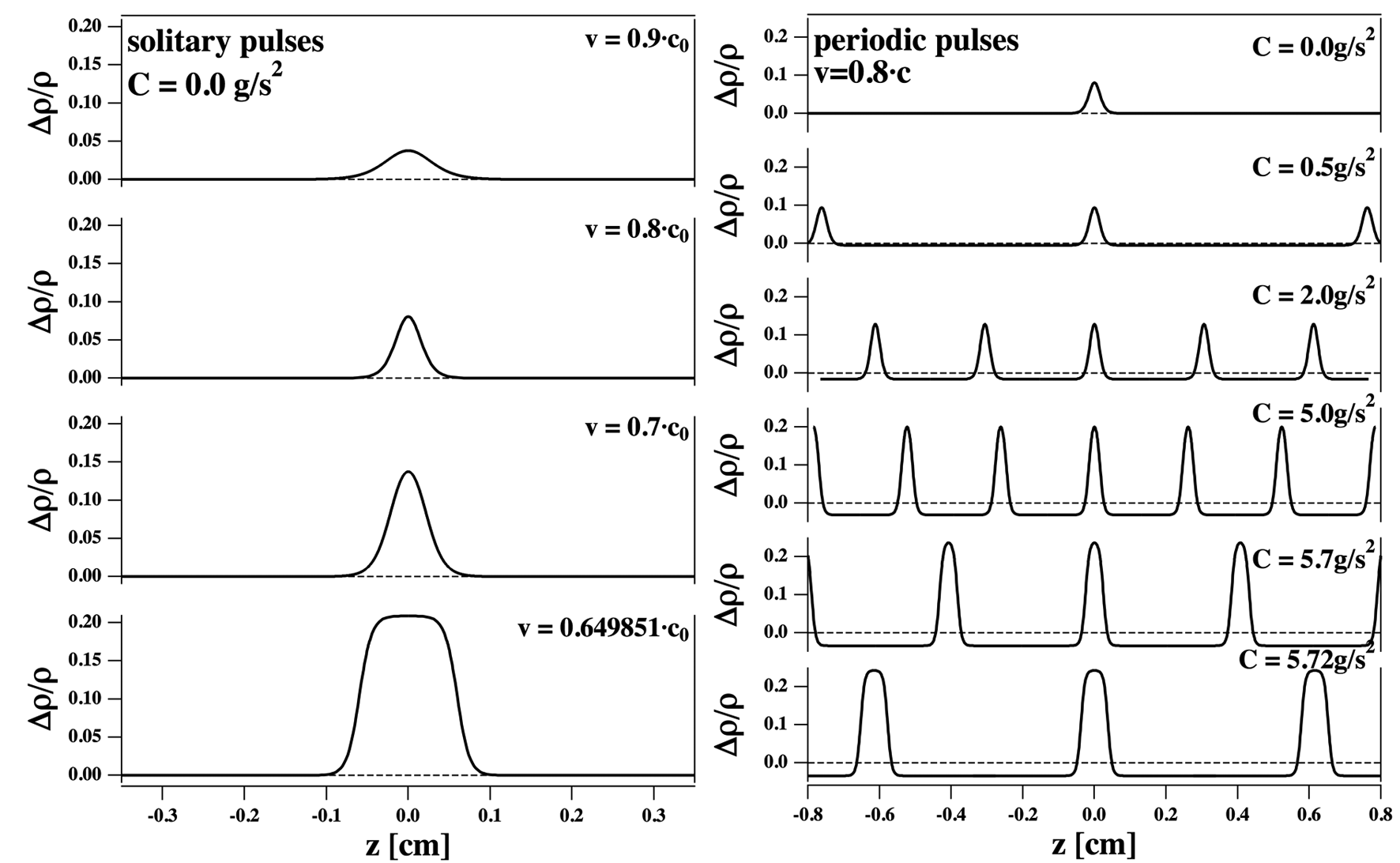}
	\parbox[c]{17cm}{ \caption{\textit{Solutions to eq.\,(\ref{eq:Theory8}) for different values of the velocity $v$and the integration constant $C$. Left: For $C=0$g/s$^2$  one obtains solitary pulses. Their peak amplitude depends on the velocity, $v$. Right: Solutions for $v=0.8\cdot c_0$ and various values of the integration constant $C$.  For all $C>0$g/s$^2$ the solutions are periodic and one finds an undershoot. The pulse distance is a function of $C$.}
	\label{Figure_Theor2}}}
    \end{center}
\end{figure*}
\section*{Results}
\subsection*{Solitary and periodic pulse solutions:}
Boundary conditions are needed to determine the desired solutions.   In the following we assume that the total length of the nerve is fixed. This implies that the mean area density of the nerve membranes stays constant (i.e., $\left<\Delta \rho^A\right>=0$), and that positive peaks must coexist with regions of negative density change. As shown below, this condition leads to periodic pulse solutions with an undershoot of the pulse (hyperpolarization) and refractory periods.  We use the elastic parameters for the lipid DPPC given above. However, as shown in \cite{Heimburg2005c} values appropriate for biological membranes would work quite as well. The value for $C$ will be varied. For instance, for a fixed propagation velocity $v=0.8 c_0$ one obtains stable solutions in the range $0$ g/s$^2\le C\le 5.72$ g/s$^2$.\\

The potential $V(\Delta \rho^A)$ is a polynomial of order 4 with negative sign in the highest order term. Therefore, this potential can display two maxima and one minimum. Stable solutions are confined to be between the two maxima; all other solutions would be unstable. For $C=0$ g/s$^2$ the potential is shown in Fig.\,\ref{Figure_Theor1}, left. It displays one local maximum at  $\Delta \rho^A=0$ and a second maximum at positive $\Delta \rho^A$ with higher value. The shape of the soliton is defined by the movement within the potential indicated by the arrow. It is defined by moving within the potential from the local maximum at $\Delta \rho^A=0$ to that positive value for $\Delta \rho^A$ for which the potential assumes the identical value. This defines the amplitude of the soliton. One obtains the isolated solitary solutions shown in Fig. \ref{Figure_Theor2}, left. For each amplitude one obtains exactly one soliton with well-defined velocity somewhat smaller than the small-amplitude speed of sound, $c_0$. Larger amplitude results in slower propagation velocity. There is a limiting velocity, $v_{limit}=c_0^2-p^2/6q\approx 0.65\cdot c_0$ at which the solitons assume maximum amplitude and minimum velocity \cite{Heimburg2005c}. The solutions consist of a local change in density above the baseline. The neuron is considered being infinitely long. Therefore the mean density is still the baseline value even if it is locally different from zero. The solitary character is a consequence of the potential having zero slope at $\Delta \rho^A=0$.  If the movement of the membrane starts from a value of $\Delta \rho^A>0$ one obtains periodic solutions. However, those solutions display a mean density change larger than zero and are therefore no solutions are possible under the assumptions of constant nerve length.

For $0 < C \le 5.72$\,g/s$^2$, the first maximum of $V$is located at a negative value of $\Delta \rho^A$. This is shown in Fig.\,\ref{Figure_Theor1}, center. Under these conditions no solitary solution can exist because that would result in $\left<\Delta \rho^A\right><0$. However, there are periodic solutions that display a mean density change of zero. Such solutions are shown in Fig.\,\ref{Figure_Theor2} (right) a fixed velocity of $v=0.8\cdot c_0$. For each value of $C$ and given velocity $v$ pulse separation and amplitude are uniquely defined.

The upper limit for the integration constant is $C\approx 5.72$ g/s$^2$. This case is shown in Fig. \ref{Figure_Theor1} (right). One recognizes that for this value of $C$ the two maxima of the potential assume the same value. This influences the shape of the pulse such that it becomes flat on the top (see Fig.\,\ref{Figure_Theor2} (right), bottom trace).  No stable solutions exist for $C > 5.72$\,g/s$^2$ and for $C< 0$\,g/s$^2$.

In summary, one finds the following:
\begin{enumerate}
\item The integration constant $C$ depends on the experimental conditions. There exist solitary and periodic solutions for the differential equation \ref{eq:Theory9} and the boundary condition of constant mean density.
\item For each $C$ and each velocity $v$ there is exactly one solution with a mean density change of zero. For $C=0$ g/s$^2$ this is a singular pulse (soliton) and for $C>0$ g/s$^2$ it is a periodic pulse.
\item There is a minimum distance between pulses dictated by the requirement that the mean density is constant, corresponding to a refractory period (Fig. \ref{Figure_Theor2}). 
\item For each pair of pulse amplitudes and pulse distances there is exactly one velocity. 
\end{enumerate}
This implies that the velocity of the pulse train depends on amplitude and frequency of the stimulation process. It further implies that any change of the amplitude resulting from a change in the conditions (e.g., the presence of anesthetics, changes in temperature or pressure, etc.) will have a predictable consequence for the propagation velocity and the spatial distance between pulses. This means that the individual pulses of the periodic solutions influence each other. One further result from Fig.\,\ref{Figure_Theor2} is that increasing pulse amplitude first leads to a narrowing of the pulses. When the pulses approach their maximum amplitude, they become broader and display a plateau at maximum density. Pulse shapes of this kind are known in nerves, for example the prolonged action potentials of squid axons that occur under perfusion with aqueous media containing unusual salt composition \cite{Kobatake1971}.

The above considerations apply to an axon that is infinitely long.  In such a system, there are no constant velocity, shape-preserving solutions involving a {\em finite\/} number of pulse peaks. Solutions in which a finite segment of a periodic solution is joined smoothly to the solution $\Delta \rho^A = 0$ is expected to be long-lived since only the limited material at the interface can lead to its disintegration.  Of course, real nerves have finite length and only a limited number of pulses can fit into a neuron at a time. It is therefore to be expected that the situation in the real nerve is close but not identical to the idealized situation described above.  In particular, we suggest that the pulse doublets observed in the locust femoral nerve (described below) represent a limiting case of the periodic solutions considered here. 
\begin{figure}[htb!]
    \begin{center}
	\includegraphics[width=8.0cm]{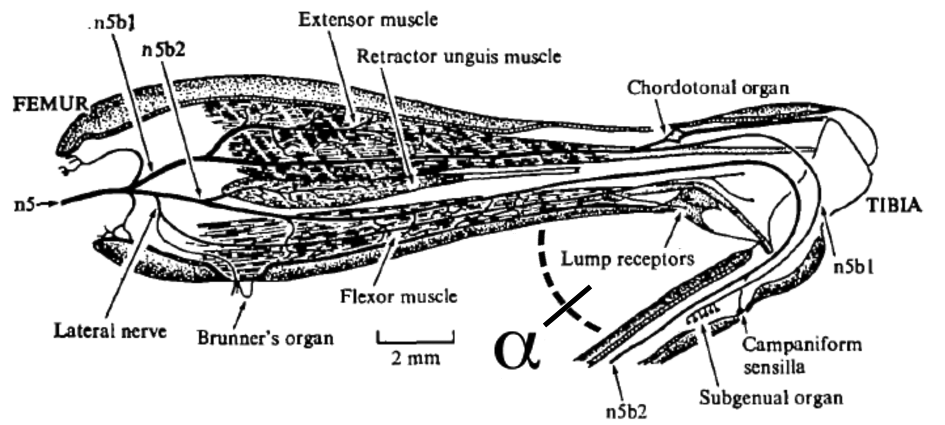}
	\parbox[c]{8cm}{ \caption{\textit{Representation of the locust metathoracic leg showing the main nerves and specifically the lateral nerve n5b2 we studied (image adapted from \cite{Heitler1977}). The femoral nerve extends along the whole leg. The angle of leg extension, $\alpha$,  influences firing frequencies (cf., Fig. \ref{Figure_Exp3}). }
	\label{Figure_Exp0}}}
    \end{center}
\end{figure}
The existence of periodic solutions, such as those shown in Fig.\,\ref{Figure_Theor2}, is a very robust result.  Arguments similar to those made here can be adopted to find periodic solutions for essentially any system exhibiting isolated solitons.  This includes, e.g., the FitzHugh-Nagumo model \cite{Fitzhugh1961, Nagumo1962} which is closely related to the Hodgkin-Huxley model.  The existence of (approximate) periodic pulses is thus not likely to be useful in discriminating between models of the action potential.  In this connection, we note that the model employed here was developed to describe myelinated nerves.  In \cite{Heimburg2005c} we suggested that myelination restricts the nerve to longitudinal density perturbations. This allowed us  to adopt the one-dimensional representation of the propagating pulse given in eq. \ref{eq:Theory8} above. Without a myelin sheet we expect that the mechanical dislocations will contain out-of-plane components which would render the pulse significantly slower. So far, we have no mathematical model for this scenario. However, we expect such pulses to be based on physics similar to the longitudinal pulses described above. Below, we compare our  theoretical results to pulse doublets in locust femoral nerves. 

\subsection*{Locust femoral nerve: Pulse trains and minimum distance doublets}
\begin{figure}[htb!]
    \begin{center}
	\includegraphics[width=8cm]{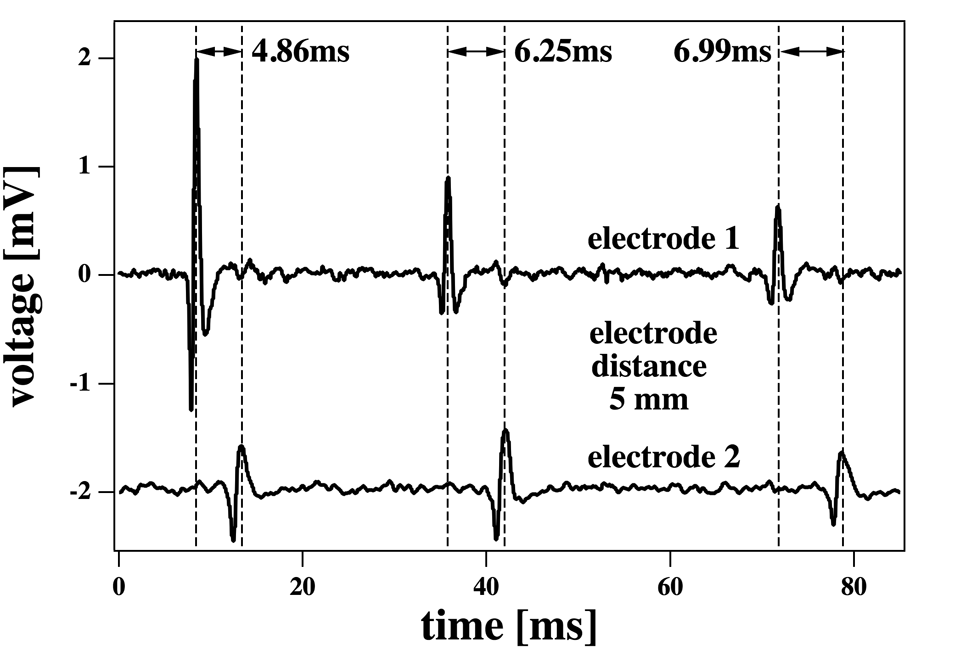}
	\parbox[c]{8cm}{ \caption{\textit{Recordings from the femoral nerve of the locust: Three different pulse originating from three different neurons that are distinctly different in amplitude and shape. Indicated are the recordings from two electrodes that are separated by 5\,mm. The time lags between the two electrodes yield the pulse propagation velocity which is of order 1\,m/s for all three neurons (see text).}
	\label{Figure_Exp1}}}
    \end{center}
\end{figure}

The femoral nerve of the locust contains three neurons (Fig.\,\ref{Figure_Exp0}) from the extension of the femoro-tibial joint. The three neurons typically display periodic patterns with frequencies and amplitudes specific for each unit. The firing frequency also depends on the angle of extension, $\alpha$ (see Fig.\,\ref{Figure_Exp0}). Therefore, in a recording one can distinguish the firing patterns of the three neurons and analyze them separately (Fig.\,\ref{Figure_Exp1}). We have labeled them by signal amplitude and refer to them as neuron 1, 2 and 3. We have measured the pulse velocity by recording the action potentials at two positions of the femoral nerve separated by 5mm (Fig.\,\ref{Figure_Exp1}). We found
\begin{enumerate}
  \item $v=1.02 \pm 0.01$ m/s for neuron 1 (large amplitude)
  \item $v=0.79 \pm 0.01$ m/s for neuron 2 (medium amplitude)
  \item $v=0.73 \pm 0.01$ m/s for neuron 3 (small amplitude)
\end{enumerate}
\begin{figure}[hb!]
    \begin{center}
	\includegraphics[width=8cm]{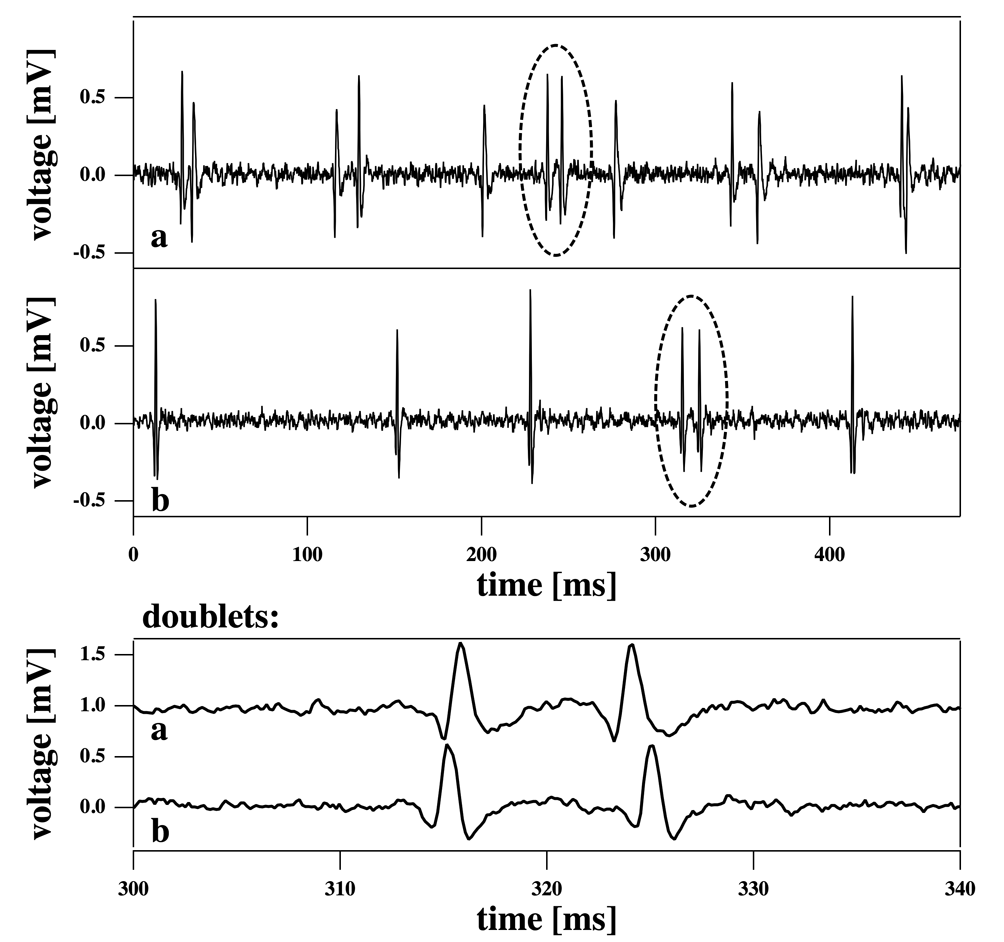}
	\parbox[c]{8cm}{ \caption{\textit{Pulse doublets occurring during the recordings of the femoral nerve. Top: Recording from neuron 2. Center: Recording from neurons 1 and 2. Bottom: The two doublets marked in the top traces given in higher resolution. }
	\label{Figure_Exp2}}}
    \end{center}
\end{figure}

Most of the pulses in the three neurons have time intervals of several times 10\,ms, depending on the angle of extension of the leg.  Fig.\,\ref{Figure_Exp2} shows recordings that display pulse doublets with identical amplitudes and separations. They occur in all three neurons. These pulse doublets occurred quite frequently and account for some 10--20\% of all action potentials. The total length of the locust femoral nerve is about 2\,cm. The pulse width is about 2\,ms. The minimum distance of pulses we have recorded is about 11\,$\pm$2 ms (see below). In 11\,ms the nerve pulse travels approximately 1\,cm. This implies that one cannot fit more than 2 or 3 pulses in the same neuron at the same time. All pulses with distances larger than 20--30\,ms must be regarded as uncorrelated, i.e., about 80-90\% of all pulses. By uncorrelated we mean that the pulse shapes of those pulses cannot influence each other because they do not coexist in a nerve. However, the minimum distance pulse doublets are simultaneously present in the neurons and can therefore display a mutual influence that may influence the separation and shape of the pulses. For this reason we will in the following consider the doublets as an approximation to the periodic solutions in the grasshopper nerve. It should be noted that we recently recorded similar pulse doublets and even triplets in the crayfish motor neuron 3 (data will be published elsewhere).
\begin{figure}[b!]
    \begin{center}
	\includegraphics[width=7cm]{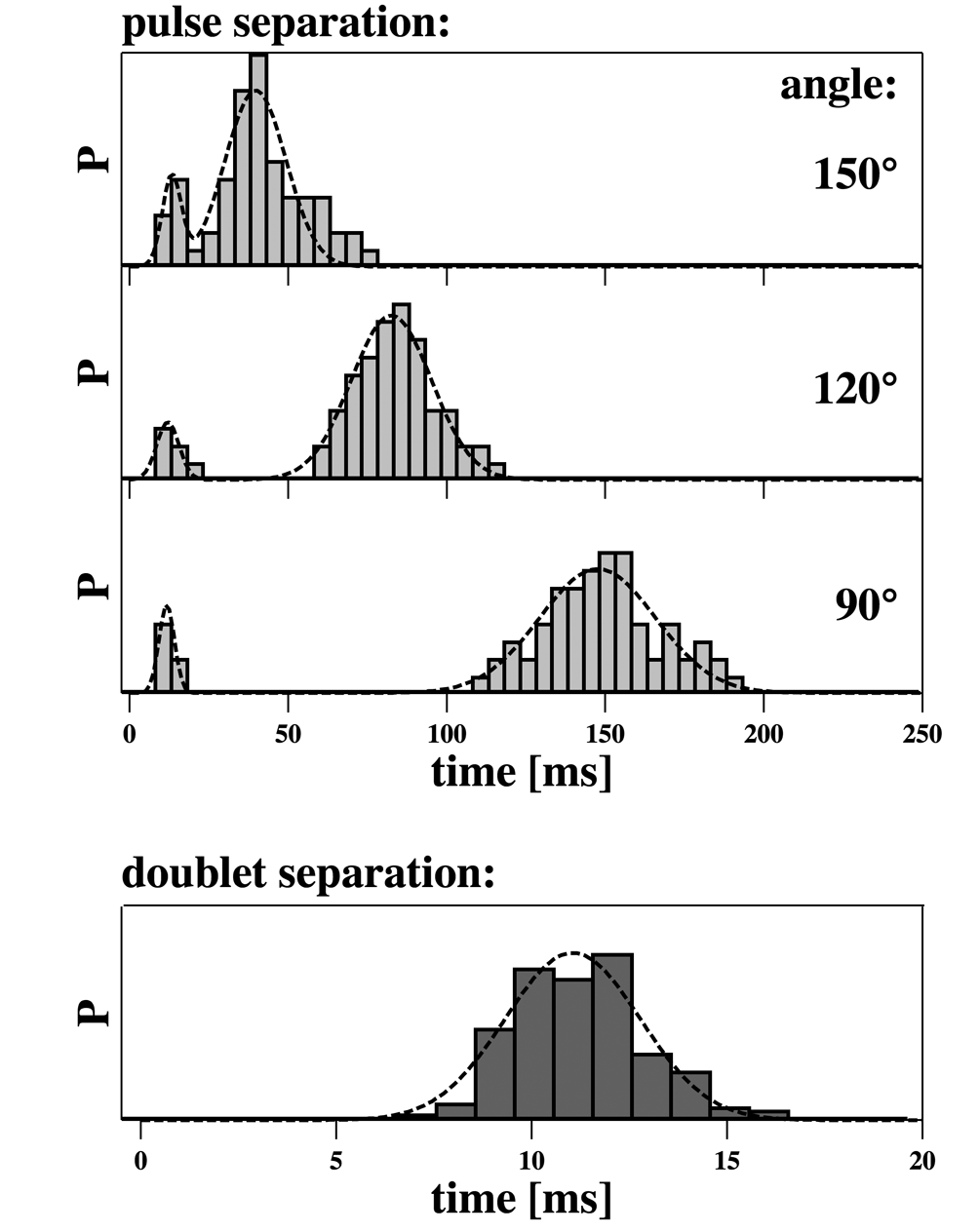}
	\parbox[c]{8cm}{ \caption{\textit{Top: Histograms of pulse separation in neuron 2 for three different angles of extension: $\alpha=$150$^{\circ}$, 120$^{\circ}$ and  90$^{\circ}$. Each histogram contains two peaks, one of which is dependent and the other one being independent of the angle of extension, $\alpha$. Bottom: The small interval peak corresponds to a doublet spacing of about 11\,ms. Details are given in table \ref{Table1}.}
	\label{Figure_Exp3}}}
    \end{center}
\end{figure}
We have recorded a histogram of pulse intervals of neuron 2 as a function of the angles of extension (Fig.\ref{Figure_Exp3}). One finds that for each of the three angles, 90$^{\circ}$, 120$^{\circ}$ and 150$^{\circ}$, the histogram displays 2 maxima. We fitted them to Gaussians (Fig.\,\ref{Figure_Exp3}) and determined the mean peak positions and the percentage of doublet intervals (shown in Table\,\ref{Table1}). About 10\% of all intervals in the recordings are related to doublets, corresponding to about 20\% of all action potentials being doublet peaks. While we find that the longer interval is strongly affected by the angle of leg extension, the short interval corresponding to the doublet peak intervals is unaffected by the angle of extension within error. For this reason they are listed independently in table \ref{Table1}. The short interval is on average around 11\,ms corresponding to the doublet spacing. The width/distance ratio for the doublet peaks is $0.1-0.2$ depending on the somewhat arbitrary criterion for defining width. The closest distance of the theoretical periodic pulses (Fig.\,\ref{Figure_Theor2}) corresponds to a width/distance ratio of about 0.125, i.e., it is of same order. Therefore we suggest that the doublets correspond to the closest possible pulses in our theory.
\begin{table}[htdp]
\caption{Mean intervals of action potentials in neuron 2 extracted from the peaks of the distributions in Fig.\,\ref{Figure_Exp3}}
\begin{center}
\begin{tabular}{|c|cc|cc|c|}
\hline 
 & \multicolumn{2}{|c|}{peak 1}  & \multicolumn{2}{|c|}{peak 2} &\% doublet \\
angle & position& width & position & width & intervals \\
 &  [ms] &  [ms] &  [ms] &  [ms] &  \\
\hline
150$^{\circ}$ & &  & 39.7 & 13.5 & 8.5 \\
120$^{\circ}$ &  & & 82.5  & 17.9 & 9.2\\
90$^{\circ}$ &  & & 147.5 & 25.8  & 13.9 \\
\hline
all  & 11.0  & 2.4 &   &  & \\
angles &    &   &   &  & \\
\hline 
\end{tabular}
\end{center}
\label{Table1}
\end{table}

\section*{Discussion}
The Hodgkin-Huxley model for the nerve pulse describes the nerve impulse as a voltage pulse generated by time- and voltage-dependent opening of ion channel proteins. As we have discussed before, this model is unable to explain a number of features of the nerve pulse including the generation of reversible heat and the observed thickness and length changes \cite{Heimburg2005c, Heimburg2007b, Heimburg2008, Andersen2009}.  Reversible heat is a particular challenge to the HH-model because it implies that the physics of the nervous impulse is based on thermodynamically reversible processes while the HH-model is exclusively based on irreversible processes due its reliance on the flow of ions along concentration gradients. In order to address these problems we have proposed that the nerve pulse instead consists of an electromechanical density pulse (soliton) that does not rely on ion channel proteins. In particular, we have shown that close to melting transitions in biomembranes the possibility of soliton propagation, i.e. of localized pulses, exists. This model automatically addresses the observed changes in length \cite{Wilke1912b} and \cite{Tasaki1989} and thickness \cite{Iwasa1980a, Iwasa1980b, Tasaki1982a, Tasaki1982b, Tasaki1989, Tasaki1990} and the reversible heat production \cite{Abbott1958, Howarth1968, Ritchie1985, Tasaki1989, Tasaki1992a}.

The key feature of our theory is a localized pulse with constant entropy, i.e., without net heat exchange with the environment. Since our theory is of thermodynamic nature, any change in a thermodynamic variable that has the potential to move the membrane through its transition is also able to generate a pulse. This includes pulse generation by changes in temperature, voltage, lateral pressure, pH, calcium concentration, etc. In particular, we have shown that this theory implies a mechanism to explain anesthesia \cite{Heimburg2007b, Heimburg2007c}.  So far, however, our theory has not included some of the features observed for the nervous impulse including hyperpolarization, refractory periods, and pulse trains.  These features are addressed in the present paper. Our original approach to pulse propagation refers to an infinitely long axon that is quasi-one-dimensional, meaning that the thickness of the nerve is negligible compared to the length of the pulse, and that the density perturbation is of longitudinal nature. The pulse consists of a locally compressed region in which the membrane is transiently pushed through its melting transition while releasing and reabsorbing  the latent heat of the transition. For an infinitely long nerve the mean area density in the presence of a singular pulse is unaffected. However, for a real nerve of finite length the overall length of the nerve should decrease while a nerve pulse is traveling along the axon. Such length changes have been observes in isolated nerves \cite{Wilke1912b, Tasaki1989}. However, it seems unlikely that a nerve within a body can undergo significant length changes because they are attached to muscles or other nerves with more or less fixed positions. Therefore, we must distinguish between nerves with constant length and nerves removed from bodies without any external constraint on length. 

Here, we have shown that the constraint of fixed overall length leads to the emergence of periodic solutions. These solutions share the property that their average area density is constant, leading to an undershoot under the baseline that is enforced by mass conservation. For each pair of pulse distances and amplitudes there is a well-defined velocity that is smaller than the speed of sound in such membranes. Further, the model implies that there is a minimum distance between pulses or a maximum frequency in a pulse train. Here, we have provided recordings from locusts (locusta migratoria) that display periodic pulses with long intervals, but with significant likelihood also minimum distance pulse doublets. In contrast to other pulses in this nerve, the doublet spacing in time is unaffected by changes of the angle of extension of the locust leg, and no pulses of closer distance were found in the individual neurons. Since the femoral nerve is only about 2\,cm long and the pulse propagation velocity is about 1\,m/s (see results section) one can conclude that action potentials with a distance larger than 20\,ms cannot be found simultaneously in a neuron. Therefore, they cannot influence each other and must be considered uncorrelated. In the femoral nerve we found to classes of pulses. One class with large time intervals indicating separations too large for temporal coexistence, and pulse doublets with spatial distances smaller than the overall length of the nerves. Such pulses can influence each other, and we consider them here as an approximation to a periodic solution for a nerve that has a length that is not larger than two pulse distances. The distance of the pulses in the femoral nerve is about 11\,ms corresponding to about 1\,cm in space. The individual pulse width is 1-2\,mm (or 1--2\,ms). Thus, the minimum pulse distance in the doublets is about 5-10 pulse widths. The minimum distance in the electromechanical model is about 8 pulse widths. It should be stressed that the theory does not allow for stable doublet solutions. However, we consider the doublets as an approximation to such solutions for the case of a nerve with a length only slightly larger than 2--3 pulse intervals. It should be added here that in our electromechanical description a simple linear relation between density and voltage has been assumed. Therefore, the integral of voltage changes over time (or space) should depend in a simple manner on the integrated density changes.

Another feature of our theory is that it attributes a meaning to myelination. We have chosen parameters leading to pulse velocities of approximately 100\,m/s, comparable to the pulse velocity of myelinated nerves and much larger than those found in the locust femoral nerve.  We have noted before that our electromechanical theory assumes longitudinal compression \cite{Heimburg2005c}, and we do not consider transverse excitations normal to the nerve membrane. However, it is well-known that longitudinal wave propagation is significantly faster than transverse propagation. E.g., the speed of sound in bulk water is 1500\,m/s while a surface wave displays a typical velocity of about 1 m/s. Therefore, it is to be expected that surface excitations in nerves involving changes in membrane curvature are considerably slower than longitudinal pulses. Thus, if one prevents out-of-plane waves by spatial confinement, one expects faster pulse propagation. This is the role of myelinization in the soliton theory. It has already been noted by Abbott et al. \cite{Abbott1958} that the magnitude of the reversible heat changes suggest that the myelinated regions of the nerve play an active role in the pulse propagation process. This is not the case in the Hodgkin-Huxley model in which myelinization plays the role of an electrical insulator preventing an active amplification of the nerve signal. Therefore, they assigned a special role to the `nodes of Ranvier' which are the only parts of the myelinated nerve that could amplify the nerve signal. In the soliton model, the `nodes of Ranvier' do not play an active role different from the rest of the nerve. The construction of a theory for transverse pulse propagation is a far more demanding task.  However, such a model would offer a better description of non-myelinated nerves. Thus, although the present results are suggestive, their relevance to the non-myelinated locust femoral nerve has not been demonstrated.

A further notable feature of the electromechanical soliton theory is that it does not require ion channel proteins. Such proteins are, however, indispensable in the Hodgkin-Huxley model. In the present model the most important requirement for obtaining localized pulses is the presence of a cooperative melting transition in the biomembrane. This transition is not only responsible for the reversible heat but also for the localization of the pulse.  Interestingly, we and others have shown that in this transition the lipid membrane displays quantized ion conduction events that are indistinguishable from the channel events typically attributed to protein ion channels, even in the complete absence of proteins.  In particular, one finds a straight-forward explanation of these quantized current events in inevitable thermodynamic fluctuations.  (For a review, see \cite{Heimburg2010}). Thus, during the pulse propagation, one expects that the lipid membrane should become permeable for ions and that quantized conduction events should occur.  Similarly, any agency that inhibits pulse formation in the soliton model would also inhibit quantized currents since the two phenomena are unavoidably coupled thermodynamically. 


\section*{Summary}
We have shown here that the soliton theory for electromechanical action potentials in nerves possesses the following features:
\begin{itemize}
  \item It generates localized voltage pulses while displaying a reversible heat and mechanical changes including shortening and swelling
  \item When the overall length of the nerve is constant, one obtains periodic solutions.
    \item One finds an undershoot or hyperpolarization.
  \item The periodic pulses display a minimum distance of about 5-10 pulse widths as a consequence of mass conservation.
  \item This pulse separation is approximately the same as that found in the femoral nerve of the locust.
  \item During the pulse one finds quantized ion conduction events through the membrane resembling those usually attributed to ion   channel proteins.
\end{itemize}
All features of the nerve pulses and the generation process display unavoidable thermodynamic couplings with a predictable influence of changes in temperature, lateral pressure, pH, calcium concentration or anesthetics. Thus, we have shown that the soliton theory for nerves is able to explain most known features of nerves, the effect of anesthetics \cite{Heimburg2007c} and the emergence of ion channel phenomena from the thermodynamics of the membrane without employing molecular features of the membrane components.

\subsection*{Acknowledgments:}
Experiments were partially carried out at the Zoology Dept. of the University of G\"ottingen, Germany, during a guest visit of E. Villagran Vargas. He acknowledges funding from the mexican National Council for Science and Technology (ConaCyt). A. Ludu was a guest of the Niels Bohr International Academy during a visit in 2009.


\small{

\begin{thebibliography}{10}

\bibitem{Heimburg2005c}
Heimburg, T., and A.~D. Jackson.
\newblock 2005.
\newblock On soliton propagation in biomembranes and nerves.
\newblock Proc.\ Natl.\ Acad.\ Sci.\ USA 102:9790--9795.

\bibitem{Heimburg2007b}
Heimburg, T., and A.~D. Jackson.
\newblock 2007.
\newblock On the action potential as a propagating density pulse and the role
  of anesthetics.
\newblock Biophys. Rev. Lett. 2:57--78.

\bibitem{Heimburg2008}
Heimburg, T., and A.~D. Jackson, 2008.
\newblock Thermodynamics of the nervous impulse.
\newblock In Structure and Dynamics of Membranous Interfaces. (Nag, K.,
  editor). Wiley, 317--339.

\bibitem{Andersen2009}
Andersen, S. S.~L., A.~D. Jackson, and T.~Heimburg.
\newblock 2009.
\newblock Towards a thermodynamic theory of nerve pulse propagation.
\newblock Progr. \ Neurobiol. 88:104--113.

\bibitem{Wilke1912a}
Wilke, E.
\newblock 1913.
\newblock Das {P}roblem der {R}eizleitung im {N}erven vom {S}tandpunkte der
  {W}ellenlehre aus betrachtet.
\newblock Pfl\"ugers Arch. 144:35--38.

\bibitem{Wilke1912b}
Wilke, E., and E.~Atzler.
\newblock 1912.
\newblock Experimentelle {B}eitr{\"a}ge zum {P}roblem der {R}eizleitung im
  {N}erven.
\newblock Pfl\"ugers Arch. 146:430--446.

\bibitem{Cole1939}
Cole, K.~S., and H.~J. Curtis.
\newblock 1939.
\newblock Electrical impedance of the squid giant axon during activity.
\newblock J.\ Gen.\ Physiol. 220:649--670.

\bibitem{Hodgkin1945}
Hodgkin, A.~L., and A.~F. Huxley.
\newblock 1945.
\newblock Resting and action potentials in single nerve fibres.
\newblock J.\ Physiol.\ London 104:176--195.

\bibitem{Kaufmann1989e}
Kaufmann, K., 1989.
\newblock Lipid membrane.
\newblock http://membranes. nbi.dk/Kaufmann/pdf/Kaufmann\_book5\_org. pdf,
  Caruaru.

\bibitem{Abbott1958}
Abbott, B.~C., A.~V. Hill, and J.~V. Howarth.
\newblock 1958.
\newblock The positive and negative heat production associated with a nerve
  impulse.
\newblock Proc. R. Soc. London. B 148:149--187.

\bibitem{Howarth1968}
Howarth, J.~V., R.~Keynes, and J.~M. Ritchie.
\newblock 1968.
\newblock The origin of the initial heat associated with a single impulse in
  mammalian non-myelinated nerve fibres.
\newblock J.\ Physiol. 194:745--793.

\bibitem{Howarth1975}
Howarth, J.
\newblock 1975.
\newblock Heat production in non-myelinated nerves.
\newblock Phil. Trans. Royal Soc. Lond. 270:425--432.

\bibitem{Ritchie1985}
Ritchie, J.~M., and R.~D. Keynes.
\newblock 1985.
\newblock The production and absorption of heat associated with electrical
  activity in nerve and electric organ.
\newblock Quart.\ Rev.\ Biophys. 392:451--476.

\bibitem{Tasaki1989}
Tasaki, I., K.~Kusano, and M.~Byrne.
\newblock 1989.
\newblock Rapid mechanical and thermal changes in the garfish olfactory nerve
  associated with a propagated impulse.
\newblock Biophys.\ J. 55:1033--1040.

\bibitem{Hodgkin1952}
Hodgkin, A.~L., and A.~F. Huxley.
\newblock 1952.
\newblock A quantitative description of membrane current and its application to
  conduction and excitation in nerve.
\newblock J.\ Physiol. 117:500--544.

\bibitem{Iwasa1980b}
Iwasa, K., I.~Tasaki, and R.~C. Gibbons.
\newblock 1980.
\newblock Swelling of nerve fibres associated with action potentials.
\newblock Science 210.

\bibitem{Heimburg1998}
Heimburg, T.
\newblock 1998.
\newblock Mechanical aspects of membrane thermodynamics. {E}stimation of the
  mechanical properties of lipid membranes close to the chain melting
  transition from calorimetry.
\newblock Biochim.\ Biophys.\ Acta 1415:147--162.

\bibitem{Halstenberg1998}
Halstenberg, S., T.~Heimburg, T.~Hianik, U.~Kaatze, and R.~Krivanek.
\newblock 1998.
\newblock Cholesterol-induced variations in the volume and enthalpy
  fluctuations of lipid bilayers.
\newblock Biophys.\ J. 75:264--271.

\bibitem{Ebel2001}
Ebel, H., P.~Grabitz, and T.~Heimburg.
\newblock 2001.
\newblock Enthalpy and volume changes in lipid membranes. i. the
  proportionality of heat and volume changes in the lipid melting transition
  and its implication for the elastic constants.
\newblock J.\ Phys.\ Chem.\ B 105:7353--7360.

\bibitem{Schrader2002}
Schrader, W., H.~Ebel, P.~Grabitz, E.~Hanke, T.~Heimburg, M.~Hoeckel, M.~Kahle,
  F.~Wente, and U.~Kaatze.
\newblock 2002.
\newblock Compressibility of lipid mixtures studied by calorimetry and
  ultrasonic velocity measurements.
\newblock J.\ Phys.\ Chem.\ B 106:6581--6586.

\bibitem{Mitaku1982}
Mitaku, S., and T.~Date.
\newblock 1982.
\newblock Anomalies of nanosecond ultrasonic relaxation in the lipid bilayer
  transition.
\newblock Biochim.\ Biophys.\ Acta 688:411--421.

\bibitem{Lautrup2005}
Lautrup, B., A.~D. Jackson, and T.~Heimburg.
\newblock 2005.
\newblock The stability of solitons in biomembranes \& nerves.
\newblock arXiv:physics/0510106 .

\bibitem{Heimburg2007c}
Heimburg, T., and A.~D. Jackson.
\newblock 2007.
\newblock The thermodynamics of general anesthesia.
\newblock Biophys.\ J. 92:3159--3165.

\bibitem{Mucke1991}
M\"ucke, A.
\newblock 1991.
\newblock Innervation pattern and sensory supli of the midleg of schistocerca
  gregaria (insecta orthopteroidea).
\newblock Zoomorph. 110:175--187.

\bibitem{Hustert1999}
Hustert, R., E.~Lodde, and W.~Gnatzy.
\newblock 1999.
\newblock Mechanosensory pegs constitute stridulatory files of grasshoppers.
\newblock J.\ Comp.\ Neurol. 410:444--456.

\bibitem{Coillot1969}
Coillot, J.~P., and J.~Boistel.
\newblock 1969.
\newblock \'{E}tude de l'activit\'e electrique propag\'ee de r\'ecepteurs \`a
  l'\'etirement de la patte m\'etathoracique du criquet, schistocerca gregaria.
\newblock J.\ Insect.\ Physiol. 15:1449--1470.

\bibitem{Remoissenet1999}
Remoissenet, M., 1999.
\newblock Waves Called Solitons.
\newblock Springer, Berlin.

\bibitem{Eichmann2002}
Eichmann, U.~A., A.~Ludu, and J.~P. Draayer.
\newblock 2002.
\newblock Analysis and classification of nonlinear dispersive evolution
  equations in the potential representation.
\newblock J.\ Phys.\ A\ - Math.\ Gen. 35:6075--6090.

\bibitem{Kobatake1971}
Kobatake, Y., I.~Tasaki, and A.~Watanabe.
\newblock 1971.
\newblock Phase transition in membrane with reference to nerve excitation.
\newblock Adv.\ Biophys. 208:1--31.

\bibitem{Fitzhugh1961}
Fitz{H}ugh, R.
\newblock 1961.
\newblock Impulses and physiological states in theoretical models of nerve
  membrane.
\newblock Biophys.\ J. 445--466.

\bibitem{Nagumo1962}
Nagumo, J., S.~Arimoto, and S.~Yoshizawa.
\newblock 1962.
\newblock An active pulse transmission line simulating nerve axon.
\newblock Proc.\ IRE 50:2061 -- 2070.

\bibitem{Iwasa1980a}
Iwasa, K., and I.~Tasaki.
\newblock 1980.
\newblock Mechanical changes in squid giant-axons associated with production of
  action potentials.
\newblock Biochem.\ Biophys.\ Research Comm. 95:1328--1331.

\bibitem{Tasaki1982a}
Tasaki, I., and P.~M. Byrne.
\newblock 1982.
\newblock Tetanic contraction of the crab nerve evoked by repetitive
  stimulation.
\newblock Biochem.\ Biophys.\ Research Comm. 106:1435--1440.

\bibitem{Tasaki1982b}
Tasaki, I., and K.~Iwasa.
\newblock 1982.
\newblock Further studies of rapid mechanical changes in squid giant axon
  associated with action potential production.
\newblock Jap.\ J.\ Physiol. 32:505--518.

\bibitem{Tasaki1990}
Tasaki, I., and M.~Byrne.
\newblock 1990.
\newblock Volume expansion of nonmyelinated nerve fibers during impulse
  conduction.
\newblock Biophys.\ J. 57:633--635.

\bibitem{Tasaki1992a}
Tasaki, I., and P.~M. Byrne.
\newblock 1992.
\newblock Heat production associated with a propagated impulse in bullfrog
  myelinated nerve fibers.
\newblock Jpn.\ J.\ Physiol. 42:805--813.

\bibitem{Heimburg2010}
Heimburg, T.
\newblock 2010.
\newblock Lipid ion channels.
\newblock Biophys.\ Chem., in print. {h}ttp://arxiv.org/pdf/1001.2524.

\bibitem{Heitler1977}
Heitler, W.~J., and M.~Burrows.
\newblock 1977.
\newblock The locust jump. {II}. {N}eural circuits of the motor programme.
\newblock J.\ Exp.\ Biol. 66:221--241.

\end{thebibliography}

}

-----------------
\end{document}